\documentclass[twocolumn,showpacs,preprintnumbers,prl,fleqn]{revtex4}
\usepackage{graphicx}
\begin{document}
\title{EPR and ferromagnetism in diluted magnetic semiconductor quantum wells}
\author{J\"urgen K\"onig$^1$ and Allan H. MacDonald$^2$}
\affiliation{
$^1$Institut f\"ur Theoretische Festk\"orperphysik, Universit\"at Karlsruhe,
76128 Karlsruhe, Germany\\
$^2$Department of Physics, University of Texas at Austin, Austin,
TX 78712}

\date{\today}

\begin{abstract}

Motivated by recent measurements of electron paramagnetic resonance (EPR)
spectra in modulation-doped CdMnTe quantum wells, [F.J. Teran {\it et al.}, 
Phys. Rev. Lett. {\bf 91}, 077201 (2003)], we develop a theory of collective
spin excitations in quasi-two-dimensional diluted magnetic semiconductors 
(DMSs).
Our theory explains the anomalously large Knight shift found in these 
experiments as a consequence of collective coupling between Mn-ion local 
moments and itinerant-electron spins.  
We use this theory to discuss the physics of ferromagnetism in (II,Mn)VI 
quantum wells, and to speculate on the temperature at which it is likely to be 
observed in n-type modulation doped systems.

\end{abstract}


\maketitle

Substitution of transition metal elements in a semiconductor lattice often  
adds local moments \cite{zunger} to the system's low-energy degrees of freedom
and can lead to qualitatively new physics.
Important progress \cite{sstreviews} has recently been achieved in 
understanding the materials science and physics of (II,Mn)VI and (III,Mn)V 
ternary compound semiconductors in which Mn has been substituted on a
relatively small fraction of the cation sites.  
When these systems are doped p-type, the Mn ions spontaneously align at low 
temperatures in both bulk and quantum-well systems with typical ferromagnetic 
transition temperatures $\sim 1{\rm K}$ in the quantum-well case and 
$\sim 100{\rm K}$ in the bulk.
Although there is broad agreement that ferromagnetism in these systems is due
to carrier-mediated interactions between Mn local moments, consensus
\cite{bookchapter} on the details of this picture is still building as the 
body of experimental studies on well characterized samples grows.  
In this connection it is intriguing that ferromagnetism has {\em never} been 
observed when these semiconductors are n-doped \cite{caveat1}.
Recent inelastic Raman and resistively detected EPR studies of n-doped 
(Cd,Mn)Te quantum wells by Teran {\it et al.} \cite{teran} provide important 
information about the role played by quantum-well electrons in correlating Mn 
ion local-moment-spin orientations.
In particular, these authors have discovered an avoided crossing between 
well-defined electron spin and Mn-ion spin resonance modes, an effect which  
demonstrates that the two subsystems can couple collectively. 

Quantum-well DMS systems are unusual in that they consist of a quasi-3D Mn ion 
system coupled to quantum-well electrons that have only two-dimensional (2D)
translational degrees of freedom.  
In this Letter we present a theoretical analysis which completely accounts for 
the observations of Teran {\it et al.}, including the size of the avoided 
crossing gap they see.  The theory sheds light on the physics that
controls ferromagnetism in DMS quantum wells and on the essential 
differences between electron- and hole-doped cases.
It follows from our theory that the gap discovered experimentally
occurs because quantum-well electrons and Mn-ion spins are coupled
ferromagnetically in n-doped systems.  According to our theory 
the ferromagnetic transition temperature is controlled by spin-orbit coupling 
strength, a quantity that can be adjusted 
{\em in situ} by biasing the quantum well \cite{rashba} and is much larger in 
p-doped systems.  We conclude that ferromagnetism {\em will}
occur in n-doped quantum wells, 
but only at temperatures well below those predicted by mean-field theory
and below those which have been studied experimentally.
In the following we first describe our theory of collective 
excitations of the coupled local-moment and quantum-well-electron spin 
subsystems, and then discuss ferromagnetism in n-doped (Cd,Mn)Te systems. 

We consider an n-type quantum well with width $d$ (in $\hat z$ direction) and 
one occupied subband. 
The quantum-well geometry makes it convenient to split the three-dimensional 
spatial coordinates into $({\bf r},z)$ where $\bf r$ is the $x$-$y$-plane 
projection.
When subband mixing is neglected \cite{leewidewells} 
the electron wavefunction separates,
$\Psi({\bf r},z) = \psi({\bf r}) \chi(z)$, where $\chi(z)$ can be chosen to be
real.  
We take advantage of the large ratio between the Mn density and the electron 
density in the samples studied by Teran {\em et al.} and in typical 
\cite{dietlhandbook94} modulation-doped (II,Mn)VI quantum-well systems, 
replacing the random distribution of Mn local moments by a continuum density 
$N_{\rm Mn}(z)$.
We are interested in the collective excitations 
in the presence of an external magnetic field ${\bf B} = (0,0,B)$.  
The Hamiltonian, $H = H_{\rm kin} + H_{\rm Zeeman} + H_{\rm sd}$ is the sum of 
three terms.
The kinetic energy of the electrons is given by
\begin{eqnarray} 
  H_{\rm kin} &=& \int d^2 r \int_0^d dz \, \chi^2(z) 
\nonumber \\ 
  && \sum_\sigma  \hat\psi^\dagger_\sigma ({\bf r}) \left(
       { (-i \hbar {\bf\nabla} + e {\bf A}/c)^2 \over 2m^*} -\mu \right) 
     \hat \psi_\sigma ({\bf r})
\end{eqnarray}
where $\bf A$ is the vector potential.
The Zeeman term reads
\begin{equation}
    H_{\rm Zeeman} = \mu_B {\bf B} \cdot \int d^2 r \int_0^d dz \left[
    g_{\rm e} \hat {\bf s} ({\bf r},z) + g_{\rm Mn} {\bf S}({\bf r},z) \right]
\end{equation}
where $\mu_B>0$ is the electron Bohr magneton.
Here,
\begin{equation}
  \hat {\bf s}({\bf r},z) = {1\over 2} \chi^2(z) \sum_{\sigma \sigma'}
        \hat \psi^\dagger_\sigma ({\bf r}) 
        {\bf \tau}_{\sigma \sigma'} \hat \psi_{\sigma'} ({\bf r}) \, ,
\end{equation}
is the electron spin density, ${\bf \tau}_{\sigma \sigma'}$ the
vector of Pauli matrices, and ${\bf S}({\bf r},z)$ is the spin density of the 
Mn system.
The ferromagnetic ($J_{\rm sd} <0$) coupling between the electron and local 
moment \cite{sstreviews} spins is described by
\begin{equation}
    H_{\rm sd} = J_{\rm sd} \int d^2 r \int_0^d dz \,
      {\bf S}({\bf r},z) \cdot \hat {\bf s}({\bf r},z) \, .
\end{equation}

As in our earlier work \cite{Koenig00} on bulk DMS ferromagnets, we develop 
our theory of elementary spin excitations in a language where the conduction 
band degrees of freedom are integrated out.  The local moments with spin
$S=5/2$ are represented by Holstein-Primakoff (HP) boson with 
retarded conduction-band-mediated interactions.
Because $g_{\rm Mn}>0$, the external magnetic field tends to align the Mn 
spins along the negative $\hat z$ direction.
For small fluctuations around this state we can approximate the spin fields 
in a coherent-state functional-integral representation of the 
HP-boson partition function by 
$S^+ \approx \bar w \sqrt{2N_{\rm Mn}(z)S}$, 
$S^- \approx w \sqrt{2N_{\rm Mn}(z)S}$, and 
$S^z = -N_{\rm Mn}(z)S + \bar w w$ where the complex variables $\bar w, w$ 
label boson coherent states.  The partition function is formally 
expressed as $Z = \int {\cal D} (\bar w w) \exp(- S_{\rm eff} [\bar w w])$
with the effective action
\begin{eqnarray}
  S_{\rm eff} [\bar w w] &=& \int_0^\beta d\tau \int d^2 r \int_0^d dz \left[
        \bar w \partial_\tau w + g_{\rm Mn}\mu_B B S^z \right] 
\nonumber \\ 
      && - \ln \det G^{-1}(\bar w w). 
\end{eqnarray}
The electron Greens function in this equation may be partitioned into 
mean-field and fluctuating terms,
 $G^{-1} (\bar ww) = (G^{\rm MF})^{-1} + \delta G^{-1}(\bar ww)$,
where 
\begin{equation}
  (G^{\rm MF})^{-1} = 
  \left( \partial_\tau - {\hbar^2 \tilde {\bf\nabla}^2 \over 2m^*} 
    -\mu \right) + {\Delta + g_{\rm e}\mu_B B\over 2}\tau^z \, ,
\end{equation}
with $\tilde {\bf\nabla} = {\bf\nabla} + (ie/\hbar c){\bf A}$, and the 
fluctuating part is
\begin{eqnarray}
   \delta G^{-1}(\bar ww) &=& -{|J_{\rm sd}|\over 2} \int_0^d dz \, \chi^2(z) 
\nonumber \\
   && \left[ \sqrt{2N_{\rm Mn}(z)S} \left( \bar w \tau^- + w \tau^+ \right) 
     + \bar ww \tau^z \right] \, .
\end{eqnarray}
The exchange contribution to the conduction band spin-splitting that appears 
in $G^{\rm MF}$ is given by $\Delta = |J_{\rm sd}| \bar N_{\rm Mn} S$ where 
$\bar N_{\rm Mn} = \int_0^d dz \chi^2(z) N_{\rm Mn}(z)$.
Expanding to second order in the boson fields and Fourier transforming we 
can write the quadratic part of the spin-wave action as
\begin{eqnarray}
  S_{\rm eff} [\bar ww] &=& {1\over \beta} \sum_m \int {d^2 k\over (2\pi)^2}
        \int_0^d dz \int_0^d dz' 
\nonumber \\ 
   && \bar w({\bf k},z, \nu_m) 
   D^{-1}({\bf k},z',z, \nu_m) w({\bf k},z', \nu_m)
\label{swaction}
\end{eqnarray}
where $\nu_m$ are the bosonic Matsubara frequencies.
The kernel $D^{-1}({\bf k},z',z, \nu_m)$ in Eq.~(\ref{swaction}) is obtained 
from a straightforward calculation which leads to a fermion loop diagram 
specified by an unwieldy expression that we do not reproduce here.  
Instead, we immediately specialize to the case of ${\bf k}=0$ probed by 
the EPR and Raman experiments of Teran {\it et al.}.
In this limit the kernel is given by the following physically transparent 
expression:
\begin{widetext}
\begin{eqnarray}
   D^{-1}({\bf k} = 0,z',z, \nu_m) &=& 
   \left[ -i \nu_m + g_{\rm Mn}\mu_B B + {|J_{\rm sd}|\over 2} 
        (n_{\downarrow}-n_{\uparrow}) \chi^2(z)\right] \delta(z-z')
\nonumber \\
   &&+ { S J_{\rm sd}^2 \chi^2(z)\chi^2(z') 
          \sqrt{N_{\rm Mn}(z)N_{\rm Mn}(z')} \over 2} \cdot
        { n_\downarrow - n_\uparrow \over i\nu_m - \Delta - g_{\rm e} \mu_B B} 
        \, ,
\label{swactionkeq0}
\end{eqnarray}
\end{widetext}
where $n_{\sigma}$ is the (2D) mean-field conduction-electron 
spin density for spin $\sigma$.  
At ${\bf k}=0$ this expression is valid at any Landau level filling factor.  
The first term on the right hand side of Eq.~(\ref{swactionkeq0})
represents the energy cost of flipping an individual Mn spin and includes the 
{\em Knight shift} $K$ contribution (see below) due to exchange-coupling 
with band electrons whose spin density depends on the position of a Mn ion 
within the quantum well.  
The second term is the correction to the energy cost (at ${\bf k}=0$) which 
occurs because the quantum-well electron system responds to Mn spin 
reorientations.
Collective excitations are determined by locating zeroes of the kernel 
determinant.
Since uniform spin orientation fluctuations correspond in our continuum theory 
to $w_0(z) \propto \sqrt{N_{\rm Mn}(z)}$, the energies $E = \hbar \Omega$
of the collective excitations we seek \cite{details} are obtained by solving 
\begin{equation}
  \int_0^d dz \int_0^d dz' w_0(z)D^{-1}(0,z',z, i\nu_m = \Omega) w_0(z') = 0 
  \, ,
\label{modes}
\end{equation}
which implies collective excitations at 
\begin{equation}
  E = {E_{\rm Mn} + E_{\rm e} \over 2} \pm
  \sqrt{ {(E_{\rm Mn} - E_{\rm e})^2 \over 4} + K \Delta} \, .
\label{solution}
\end{equation}
The quantities in Eq.~(\ref{solution}) all have simple physical 
interpretations.
The expressions $E_{\rm Mn}$ and $E_{\rm e}$ denote mean-field transition energies 
in which the magnetic quantum number is increased by one,
for Mn and electron spins respectively. 
Each mean-field excitation energy has two contributions, the Zeeman energy 
and the exchange energy due to 
coupling between the two spin subsystems.
For the Mn spin, $E_{\rm Mn} = E^Z_{\rm Mn} + K$ where the Zeeman term is
$E^Z_{\rm Mn} = g_{\rm Mn}\mu_BB$ and the {\em Knight shift} is 
\begin{equation}
   K = {  |J_{\rm sd}| (n_{\downarrow}-n_{\uparrow})  \over 2} \cdot
   {\bar N_{\rm Mn} \over N_{\rm Mn}d} \, .
\label{knight}
\end{equation}
$K/(g_{\rm Mn}\mu_B)$ is the mean exchange field experienced by the 
local moments because of the electron spin polarization.
It is analogous to the Knight shift experienced by nuclei in a spin-polarized 
electron gas in NMR experiments.  
The average Mn density in this expression, $N_{\rm Mn}$, defined 
by $N_{\rm Mn} d = \int_0^d dz N_{\rm Mn} (z)$,  
is not equal to the quantity $\bar N_{\rm Mn}$ which appeared previously in 
the mean-field quantum-well electron spin gap; 
the effective 3D electron density in Eq.~(\ref{knight})
involves the quantum-well width and the Mn distribution within 
the quantum well in a non-obvious way that follows from our analysis. 

For the quantum-well electrons, $E_{\rm e} = E^Z_{\rm e} + \Delta$, the
Zeeman term is $E^Z_{\rm e} = g_{\rm e}\mu_BB$.
Following the NMR analogy, the mean-field electron spin-splitting 
$\Delta = |J_{\rm sd}|\bar N_{\rm Mn} S$ corresponds to the 
nuclear-polarization induced Overhauser shift in electron spin-resonance 
frequencies.
Because $g_{\rm Mn}>0$, and therefore $E^Z_{\rm Mn} > 0$, the mean-field 
Mn-spin configuration points in the negative $\hat z$ direction. 
For the conduction-band electrons, there is a competition between the 
Overhauser energy $\Delta > 0$ and the Zeeman term $E^Z_{\rm e} < 0$ 
($g_{\rm e} < 0$).
The Overhauser shift dominates except at very strong magnetic fields, i.e., 
the mean-field electron spin polarization is along the negative $z$ axis.

\begin{figure}
\centerline{\includegraphics[width=7.8cm]{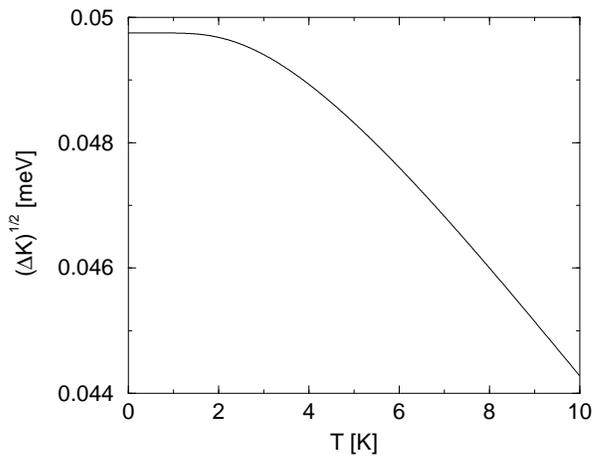}}
\caption{Temperature dependence of the gap $\sqrt{\Delta K}$.}
\label{fig1}
\end{figure}

Eq.~(\ref{solution}) is able to account quantitatively 
for the experiments of Teran {\it et al.}.
In their sample the avoided crossing is seen in a strong magnetic field near 
Landau level filling factor $\nu=3$, where the electronic state has two 
occupied majority-spin Landau levels and one minority spin, {\it i.e.}, 
$n_{\downarrow}-n_{\uparrow} = n/3 \approx 2 \times 10^{11} {\rm cm}^{-2}$.
For a Mn fraction that is constant across the quantum well we obtain 
$N_{\rm Mn} = \bar N_{\rm Mn} \approx 4.4 \times 10^{19} {\rm cm}^{-3}$ for the
studied sample \cite{teran}.
The width of the quantum well is $d \approx 10 {\rm nm}$.
At low temperatures, the Mn spins are fully polarized by the external magnetic 
field.
Using the experimental \cite{teran} low-temperature value for 
$\Delta \approx 1.65 \, {\rm meV}$, we conclude that $|J_{\rm sd}|$ in 
(Cd,Mn)Te is $0.015 \, {\rm eV \, nm}^3$, in agreement with 
Ref.~\onlinecite{furdyna}.
Using these values we conclude that $K \approx 1.5 \, {\rm \mu eV}$ and hence 
the avoided crossing gap $\sqrt{K\Delta} \approx 0.05 \, {\rm meV}$, in close 
agreement with the experimental estimate $ \approx 0.03 \, {\rm meV}$.
In the quantum Hall regime, the gap's temperature dependence may be attributed 
to the temperature-dependent Mn spin-polarization and the consequent 
temperature dependence of $\Delta$ (see Fig.~\ref{fig1}).
At these strong fields the Mn spins polarization is described by the Brillouin 
function so that 
$\Delta (T) / \Delta(0) = B_{5/2}(5g_{\rm Mn}\mu_B B / 2k_B(T+T_{\rm AF}))$ 
where the phenomenological parameter $T_{\rm AF} = 0.18 K$ accounts for
the direct antiferromagnetic Mn-Mn coupling \cite{teran}.

The close agreement between experimentally observed and calculated anticrossing
gaps supports our effective model for DMSs in which the low-energy degrees of 
freedom are local $S=5/2$ Mn spins that are exchange coupled
to band-electron spins.  
The avoided crossing establishes that both Mn and band spins behave 
collectively, and that they are coupled, conditions under which ferromagnetism
is expected.   
The ferromagnetic transition temperature of this quantum-well system 
can be estimated by applying the mean-field approach that appears to be 
generally successful when applied \cite{tcmft} to bulk (III,Mn)V materials.
Assuming uniform Mn doping and 
particle-in-a-box subband wavefunctions, the mean-field critical temperature 
for quantum wells \cite{tcqw} can be written as 
\begin{equation}
  k_B T_c^{\rm MF} = \frac{S + 1}{4} \cdot \frac{K^{\rm max} \Delta}
  {\epsilon_{\rm F}}
\label{tcmf}
\end{equation}
where $K^{\rm max} = n|J_{\rm sd}|/(2d)$ is the value of the `Knight shift' 
when the quantum-well electrons are completely spin polarized and 
$\epsilon_{\rm F}$ is the paramagnetic 2D electron gas Fermi 
energy.
Using experimental values for $\Delta \approx 1.65 {\rm meV}$,
$K^{\rm max} \approx 4.5 {\rm \mu eV}$, and the effective mass 
$m^* = 0.107 m_{\rm e}$ we find that $T_c^{\rm MF} \approx 5.6 {\rm mK}$.
  
In three-dimensional DMS ferromagnets, we expect \cite{schliemann} mean-field 
theory to be reliable when the carrier density is smaller than the Mn 
density and $\Delta < \epsilon_{\rm F}$.  
These conditions are both satisfied here.
In the quasi-2D case, however, long-wavelength collective excitations 
have a significant negative impact on tendencies toward long-range magnetic 
order.
In fact, for the model studied here which has no spin-orbit interactions, 
continuous spin-rotational invariance implies that long-range magnetic order 
is impossible at finite temperatures in 2D \cite{mermin-wagner}.  
Thermal and quantum fluctuations at low temperatures in these quasi-2D 
ferromagnets are even more important than usual because 2D electron gas 
properties lead to vanishing spin stiffness \cite{kossacki00,details}.
Long-range magnetic order is possible in (II,Mn)VI quantum wells only 
because spin-orbit coupling favors magnetization orientations perpendicular to
the quantum wells.

In symmetric quantum wells spin-orbit interactions are described by the
Dresselhaus Hamiltonian, $H_D = \gamma (-\sigma_x k_x + \sigma_y k_y) 
\langle k_z^2 \rangle$.  
Evaluating the magnetization-orientation dependence of the energy correction 
due to this term by second-order perturbation theory we find that the easy 
axis is perpendicular to the quantum well.  
The collective excitation energy gap is twice the anisotropy energy per Mn 
spin.  
We find that 
\begin{equation}
  E_{\rm so} = {\gamma^2 \langle k_z^2 \rangle^2 k_{\rm F}^2 n \over S 
    N_{\rm Mn} d \,\, {\rm max} \{ \Delta, 2\epsilon_{\rm F} \} },
\label{sogap}
\end{equation}
taking $\langle k_z^2 \rangle = (\pi/d)^2$ for the lowest quantum-well subband,
and $k_{\rm F}^2 = 2m^*\epsilon_{\rm F}/\hbar^2$.
When $E_{\rm so}$ is small the Curie temperature will be limited by collective 
fluctuations \cite{schliemann}.  Taking account of the 
expected dispersionless 2D spin-wave bands and estimating the Curie temperature
as the temperature at which $S$ magnons per Mn spin are thermally excited,   
we predict that $k_BT_c^{\rm coll} \approx (S+1/2) E_{\rm so}$.
Plugging in $\gamma \approx 12 \, {\rm meV} \, {\rm nm}^3$ for the Dresselhaus 
coefficient in the CdTe conduction band \cite{cardona}, this implies that 
$T_c^{\rm coll} \approx 0.4 {\rm mK}$ in the sample studied by Teran 
{\it et al.}, considerably smaller than the mean-field estimate.
It is interesting to note that with increasing Mn-doping concentration 
$N_{\rm Mn}$ the mean-field Curie temperature increases [Eq.~(\ref{tcmf})]
while collective-fluctuation energy decreases [Eq.~(\ref{sogap})].
For asymmetric quantum wells the spin-orbit interactions are dominated by 
Rashba coupling, and, depending on the degree of asymmetry, the
collective-excitation energy gap will become larger until, eventually, the 
system can again reach the mean-field regime.  Stronger spin-orbit
interactions in the valence band \cite{holerefs} may explain the apparent 
success of mean-field theory in predicting ferromagnetic transition 
temperatures there.  The strong antiferromagnetic interactions that 
occur between neighbouring Mn ions will reduce \cite{holerefs,sstreviews}
the number of Mn spins that can be spontaneously aligned by
the relatively weak but longer ranged carrier mediated interactions,
but should not preclude ferromagnetism when the Mn mole fraction is low.

In conclusion, we presented a theory of collective spin excitations in DMS
quantum wells which explains the anomalously large Knight shift observed
in recent experiments.  We use our theory to discuss the physics of 
ferromagnetism in n-type DMS quantum-well systems, which has not yet been 
observed.  We point out that spin-orbit interactions are necessary to support 
ferromagnetism, predict ferromagnetic transition temperatures in symmetric 
quantum wells that can be well below those implied by mean-field theory, and 
suggest that a crossover between collective-fluctuation-limited and  
mean-field-interaction-limited ferromagnetism can be observed in these systems
by using a bias voltage to adjust the spin-orbit interaction strength.  

We acknowledge helpful discussions with T. Dietl, D. Frustaglia and 
M. Governale, and thank F.J. Teran and M. Potemski for showing us their 
results prior to publication.
This work was supported by the Deutsche Forschungsgemeinschaft under the Center
for Functional Nanostructures and the Emmy-Noether program, by the Department 
of Energy under grant DE-FG03-02ER45958 and the Welch Foundation.


\begin{thebibliography}{99}
\bibitem{caveat1} 
Since Mn is an acceptor in (III,V) semiconductors, n-doping is possible only 
in (II,VI) semiconductors.
\end{thebibliography}
\end{document}